\documentclass{llncs}

\usepackage{makeidx}
\usepackage{url}
\usepackage{alltt}

\usepackage{algorithmicx}
\usepackage{algorithm}
\usepackage{algpseudocode}
\usepackage{numprint}

\begin{document}

\title{Comparing DNA sequence collections by direct comparison of compressed text indexes}
\author{Anthony J. Cox\inst{1}, Tobias Jakobi\inst{2}, Giovanna Rosone\inst{3} \\ and Ole B. Schulz-Trieglaff\inst{1}}
\institute{Illumina Cambridge Ltd., United Kingdom\\
\email{\{acox,oschulz-trieglaff\}@illumina.com}\\
\and
Computational Genomics, CeBiTec, Bielefeld University, Germany\\
\email{tjakobi@cebitec.uni-bielefeld.de}
\and
University of Palermo, Dipartimento di Matematica e Informatica, Italy
\email{giovanna@math.unipa.it}
}

\maketitle

\begin{abstract}

Popular sequence alignment tools such as BWA convert a reference genome to an indexing data structure based on the Burrows-Wheeler Transform (BWT), from which matches to individual query sequences can be rapidly determined. However the utility of also indexing the query sequences themselves remains relatively unexplored. 

Here we show that an all-against-all comparison of two sequence collections can be computed from the BWT of each collection with the BWTs held entirely in external memory, i.e. on disk and not in RAM. 
As an application of this technique, we show that BWTs of transcriptomic and genomic reads can be compared to obtain reference-free predictions of splice junctions that have high overlap with results from more standard reference-based methods.

Code to construct and compare the BWT of large genomic data sets is available at \url{http://beetl.github.com/BEETL/} as part of the \texttt{BEETL} library.

\end{abstract}

\def\bwtS#1{\textsf{bwt}_{#1}(\mathcal{S})}
\def\lcpS#1{\textsf{lcp}_{#1}(\mathcal{S})} %
\def\bwtw#1{\textsf{bwt}_{#1}({w})}
\def\lcpw#1{\textsf{LCP}_{#1}({w})}
\def\bigS{\mathcal{S}}
\def\BCR{\texttt{BCR} }
\def\BWT{\textsf{BWT}} %
\def\LCP{\textsf{LCP}} %
\def\SA{{\textsf SA}} %
\def\BCRLCP{\texttt{extLCP} }
\def\BWTE{\texttt{bwte} }
\def\sort#1{\textrm{sort}(#1)}
\def\BEETL{\texttt{BEETL}}
\def\bzip2{\texttt{bzip2}}

\def\gzip{\texttt{gzip}}
\def\7zip{\texttt{7-Zip}}
\def\BCRext{\texttt{BCRext} }
\def\recoil{\texttt{ReCoil}}

\section{Introduction}

In computer science, a \emph{suffix tree} is the classical example of an \emph{indexing} data structure which, when built from some text $T$, allows the presence or absence of a query string $S$ in $T$ to be rapidly determined. A suffix tree is several times larger than the text it indexes, but research since 2000 (well summarized in \cite{NavarroMakinen2007}) has led to \emph{compressed full-text indexes} that provide the same functionality as the suffix tree while taking up less space than the text itself. 

Of these, the \emph{FM-index} has become central to bioinformatics as the computational heart of popular sequence alignment tools such as BWA \cite{bwa2009}, Bowtie \cite{bowtie2009} and SOAP2 \cite{soap2_2009}. All these programs work in a similar way, with individual query sequences being searched for one-by-one in the index of a reference genome. The FM-index of, say, the latest human reference sequence can be viewed as a constant and precomputed, so the cost of building it is not important for this particular use case.

Constructing the FM-index of $T$ is dominated by the computation of the \emph{Burrows-Wheeler transform}, a permutation of the symbols of $T$ that also has widespread applications in data compression \cite{bookBWTAdjeroh:2008}. For large $T$, this calculation requires either a large amount of RAM or a cumbersome divide-and-conquer strategy. However, in \cite{BauerCoxRosoneCPM11,BauerCoxRosoneTCS2012}, two of the present authors demonstrated that if $T$ can be considered to be a large number of independent short patterns then its BWT can be built partially or entirely in external memory (that is, by sequential access to files held on disk). This leads us to the aim of the present work, which is to introduce some of the additional possibilities that arise if the set of query sequences is also indexed.

The search for a single pattern in an FM-index may potentially need access to any part of the BWT, requiring the entire BWT to be held in RAM to guarantee that this can be efficiently achieved. However, building on our previous work, we show that the computations needed for an all-against-all comparison of the sequences in two collections can be arranged so that the BWTs of the collections are both accessed in a series of sequential passes, permitting the comparison to be done efficiently with the BWTs held on disk. In $k$ passes, this procedure traverses all $k$-mers that are present in one or both of the two indexes. This traversal can be viewed as a template upon which different sequence comparison tasks can be defined by specifying particular sets of behaviours according to whether each of the $k$-mers encountered is unique to one or the other dataset, or shared by both. To illustrate, we show how this template may be adapted to the task of comparing transcriptomic and genomic reads from an individual eukaryotic organism to deduce exon-exon splice junctions.

In a eukaryotic genome, large tracts of intragenic and intronic DNA will be represented in the genome but not the transcriptome, but the sequences present in the transcriptome alone are far fewer and of more interest: notwithstanding experimental artefacts and the relatively rare phenomenon of RNA editing, these must span splice junctions between exons. By obtaining the genomic and transcriptomic samples from the same individual, we eliminate the possibility that differences between the datasets are due to genetic variation between individuals.

 We apply our methods to data from the Tasmanian Devil (\textit{S. Harrissii}). The hypothesis- and reference-free nature of our procedure is advantageous for \emph{de novo} projects where no reference sequence is available or cases where, as here, the reference genome is of draft quality.

\section{Methods}
\label{sec:methods}

\subsection{Definitions}
Consider a string $s$ comprising $k$ symbols from an alphabet $\Sigma =\{c_1, c_2, \ldots, c_\sigma\}$ whose members satisfy $c_1 < c_2 < \cdots < c_\sigma$. We mark the end of $s$ by appending a special \emph{end marker} symbol $\$$ that satisfies $\$ < c_1$. We can build $k+1$ distinct \emph{suffixes} from $s$ by starting at different symbols of the string and continuing rightwards until we reach $\$$. If we imagine placing these suffixes in alphabetical order, then the \emph{Burrows-Wheeler transform} \cite{bwt94,bookBWTAdjeroh:2008}  of $s$ can be defined such that the $i$-th element of the BWT is the symbol in $s$ that precedes the first symbol of the $i$-th member of this ordered list of suffixes. Each symbol in the BWT therefore has an \emph{associated suffix} in the string.  A simple way (but not the only way - see \cite{MantaciRRS07}) to generalize the notion of the BWT to a collection of $m$ strings $S=\{s_1, \ldots, s_m\}$ is to  imagine that all members $s_i$ of the collection are terminated by distinct end markers $\$_i$ such that $\$_1 < \cdots < \$_m < c_1$.

The characters of $\mathrm{BWT}(S)$ whose associated suffixes start with some string $Q$ form a single contiguous substring of $\mathrm{BWT}(S)$. We call this the \emph{$Q$-interval} of $\mathrm{BWT}(S)$ and express it as a pair of coordinates $[b_Q,e_Q)$, where $b_Q$ is the position of the first character of this substring and $e_Q$ is the position of the first character after it. This definition is closely related to the \emph{lcp-interval} introduced in \cite{Abouelhoda2004}: the $Q$-interval is an lcp-interval of length $|Q|$. If $Q$ is not a substring of any member of $S$ then a consistent definition of its $Q$-interval is $[b_Q,b_Q)$, where $b_Q$ is the position $Q$ would take if it and the suffixes of $S$ were to be arranged in alphabetical order. 

Each occurrence of some character $c$ in the Q-interval corresponds to an occurrence of the string $cQ$ in $S$. We call $cQ$ a \emph{backward extension} of $Q$. If all characters in the $Q$-interval are the same then $Q$ has a \emph{unique backward extension}, which is equivalent to saying that all occurrences of $Q$ in $S$ are preceded by $c$. %
A $Q$-interval of size 1 is a special case of unique backward extension that corresponds to there being a unique occurrence of $Q$ in $S$: we call this a \emph{singleton backward extension}.
Similarly, we can say that appending a character $c$ to $Q$ to give $Qc$ creates a \emph{forward extension} of $Q$. Since all suffixes that start with $Qc$ must also start with $Q$, the $Qc$-interval is clearly a subinterval of the $Q$-interval.

\subsection{All-against-all backward search}
We can use $Q$-intervals to describe the \emph{backward search} algorithm for querying  $\mathrm{BWT}(S)$ to compute $\mathrm{occ}(P)$, the number of occurrences of some query string $P=p_1\cdots p_n$ in $S$. This proceeds in at most $n$ stages. At stage $j$, let $Q$ be the \emph{$j$-suffix} (\emph{i.e.} the last $j$ symbols) of $P$, let $c$ be the character preceding $Q$ in $P$ and assume we know the position of the (non-empty) $Q$-interval in  $\mathrm{BWT}(S)$. The number of occurrences of $cQ$ in $S$ is given by the number of occurrences of $c$ in the $Q$-interval (see \cite{Ferragina:2000}). If this is zero, then we know that $cQ$ does not occur in $S$ and so $\mathrm{occ}(P)$ must be zero. Otherwise, we observe that the number of occurrences of $cQ$ in $S$ is by definition the size of the $cQ$-interval. The start of the $cQ$-interval is given by the count of $c$ characters that precede the start of the $Q$-interval. These are the two pieces of information we need to specify the $cQ$-interval that we need for the next iteration. At the last stage, the count of $p_1$ characters in the $p_2\cdots p_n$-interval gives $\mathrm{occ}(P)$. 

Running this procedure to completion for a single query $P$ entails counting symbols in intervals that can potentially lie anywhere in $\mathrm{BWT}(S)$, the whole of which must therefore reside in RAM if we are to guarantee this can be done efficiently. However, we show that $\mathrm{occ}(P)$ may be computed for all strings $P$ of length $k$ or less in $S$ by making $k$ sequential passes through $\mathrm{BWT}(S)$, allowing the processing to be done efficiently with $\mathrm{BWT}(S)$ held on disk. 

At the start of iteration $j$, we open an array $F$ of $\sigma$ write-only files and set each entry of an array $\Pi$ of $\sigma$ counters to zero. During the iteration, we apply the \texttt{processInterval()} function described in Figure~\ref{Alg1} to the $Q$-intervals of all $j$-suffixes of $S$ which, by induction, we assume are available in lexicographic order. With this ordering, the intervals $[b_Q,e_Q)$, $[b_{Q'},e_{Q'})$ of two consecutive $j$-suffixes $Q$, $Q'$ satisfy $e_Q \leq b_{Q'}$, which means that we can update the counters $\Pi$ and $\pi$ needed by \texttt{processInterval()} by reading the symbols of $\mathrm{BWT}(S)$ consecutively.

These arrays simulate the $\mathrm{rank}()$ function used in the FM-index  - $\Pi[i]$ holds $\mathrm{rank}(c_i,b_Q)$, the number of occurrences of $c_i$ prior to the start of the $b_Q$, whereas $\pi[i]$ counts the occurrences of $c_i$ in the $Q$-interval $[b_Q,e_Q)$ itself. The pair $(\Pi[i], \pi[i])$ that is appended to the file $F[i]$ specifies the start position and size of the $c_iQ$ interval. The last act of \texttt{processInterval()} is to update $\Pi$ to count the occurrences of each symbol up to position $e_Q$ and return the updated array ready to be passed in at the next call to the function.

At the end of the iteration, each file $F[i]$ contains the $Q$-intervals of all $(j+1)$-suffixes that start with symbols $c_i$, in lexicographic order. If we consider the files in the order $F[1],F[2],\ldots,F[\sigma]$ and read the contents of each sequentially, we have the lexicographic ordering of the $(j+1)$-suffixes that we need for the next iteration.

\begin{figure}

\begin{algorithmic}
\Function{processInterval}{$[b_Q,e_Q)$,$B$,$\Pi$,$F$}
\State Update $\Pi$ if necessary so that each $\Pi[i]$ counts occurrences of $c_i$ in $B[O,b_Q)$
\State Create $\pi$ such that each $\pi[i]$ counts occurrences of $c_i$ in $B[b_Q,e_Q)$
\For{$i=1 \to \sigma$}
\If{$\pi[i]>0$} 
\State Write $[\Pi[i],\pi[i])$ to file $F[i]$
\EndIf
\EndFor
\State $\Pi \gets \Pi +\pi$
\State \Return $\Pi$
\EndFunction
\end{algorithmic}
\caption{Given a $Q$-interval $[b_Q,e_Q)$ of $Q$ in a BWT string $B$, the function \texttt{processInterval()} computes the $c_iQ$-intervals for all backward extensions $c_iQ$ of $Q$ that are present in $[b_Q,e_Q)$ and appends them to the appropriate file $F[i]$ ready for processing during the next iteration.}
\label{Alg1}
\end{figure}

\begin{figure}

\begin{algorithmic}
\While{(1)}
\While{($\mathrm{gotQ}==\texttt{true}$)}
\State $\mathrm{gotQ}=\mathrm{getNextInterval}(b_Q,e_Q,m_Q,\mathrm{BWT}(S_A))$
\If{ ($\mathrm{gotQ}==\texttt{false}$) or ($m_Q==\texttt{true}$)} \State break \EndIf
\State doAOnlyBehaviour()
\State processInterval$(b_Q,e_Q,\mathrm{BWT}(S_A),\Pi_A,F_A)$
\For{$i=1 \to \sigma$}
\If{($pi_A[i]>0$)} 
\State Write \texttt{false} to file $M_A[i]$
\EndIf
\EndFor
\EndWhile
\While{($\mathrm{gotR}==\texttt{true}$)}
\State $\mathrm{gotR}=\mathrm{getNextInterval}(b_R,e_R,m_R,\mathrm{BWT}(S_B))$
\If{ ($\mathrm{gotR}==\texttt{false}$) or ($m_R==\texttt{true}$)} \State break \EndIf
\State doBOnlyBehaviour()
\State processInterval$(b_R,e_R,\mathrm{BWT}(S_B),\Pi_B,F_B)$
\For{$i=1 \to \sigma$}
\If{($pi_B[i]>0$)} 
\State Write \texttt{false} to file $M_B[i]$
\EndIf
\EndFor
\EndWhile
\If{($\mathrm{gotQ}==\texttt{false}$)} \State break \EndIf
\State doSharedBehaviour()
\State processInterval$(b_Q,e_Q,\mathrm{BWT}(S_A),\Pi_A,F_A)$
\State processInterval$(b_R,e_R,\mathrm{BWT}(S_B),\Pi_B,F_B)$
\For{$i=1 \to \sigma$}
\If{($pi_A[i]>0$) and ($pi_B[i]>0$)} 
\State Write \texttt{true} to files $M_A[i]$, $M_B[i]$
\ElsIf{($pi_A[i]>0$)} 
\State Write \texttt{false} to file $M_A[i]$
\ElsIf{($pi_B[i]>0$)} 
\State Write \texttt{false} to file $M_B[i]$
\EndIf
\EndFor
\EndWhile
\end{algorithmic}

\caption{Pseudocode for stage $j$ of the all-against-all comparison of the BWTs of the collections $S_A$ and $S_B$. Consecutive calls to function $\texttt{getNextInterval()}$ are assumed to populate $b_Q$, $e_Q$ and $m_Q$ with details of the $Q$-intervals of the $j$-suffixes of the relevant BWT in lexicographic order, returning \texttt{false} once the list of intervals has been exhausted. In practice, these intervals are read sequentially from the sets of files $F_A$, $F_B$, $M_A$ and $M_B$ that were generated during the previous execution of this procedure.}
\label{Alg2}

\end{figure}

\subsection{All-against-all comparison of two BWTs}

The concept of all-against-all backward search can be extended to compute the union of all suffixes of length $k$ or less present in two collections $S_A$ and $S_B$ by making $k$ passes through their BWTs.  Figure~\ref{Alg2} describes the logic of a single pass. Conceptually, each pass is a simple merge of the two lists of $Q$-intervals of a given length, with the ordering of $Q$-intervals being determined by the lexicographic ordering of their associated suffixes $Q$. The notable implementation detail is that associating an additional bit $m_Q$ with each $Q$-interval avoids the need to store and compare strings when deciding whether $Q$-intervals from the two collections are associated with the same suffix. As in the previous section, this insight is best understood inductively: if a $Q$-interval is shared (which we know because $m_Q$ is set to \texttt{true}), then any common backward extensions $cQ$ must also be common to both collections (and $m_{cQ}$ must be set to \texttt{true} to reflect that).

Each suffix $Q$ we encounter during the execution of the algorithm in Figure~\ref{Alg2} is either present  in $S_A$ only, present in $S_B$ only, or common to both $S_A$ and $S_B$ and the algorithm calls different functions in the event of these three possibilities. We show how to specify the behaviour of these three functions so as to compute differences between genomic and transcriptomic sequence data from an individual eukaryotic organism.

\begin{figure}[!htb]
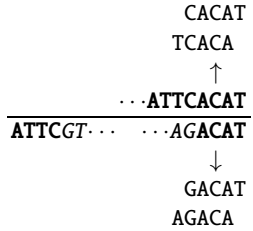


\begin{tabular}{ccc}
\qquad\qquad\qquad\qquad \qquad
&
\begin{tabular}{r}
\texttt{\ CACAT} \\
\texttt{TCACA\ } \\
$\uparrow$\texttt{\ \ \ }\\
\texttt{$\cdots$\textbf{ATTCACAT}}\\
\hline
\texttt{\textbf{ATTC}\textit{GT}$\cdots$\ \ $\cdots$\textit{AG}\textbf{ACAT}} \\
$\downarrow$\texttt{\ \ \ }\\
\texttt{GACAT} \\
\texttt{AGACA\ } \\
\end{tabular} 
&
\qquad \\
\end{tabular}

\caption{
Simple example of genome/transcriptome comparison. In the genome (below the line), the exons \texttt{ATTC} and \texttt{ACAT} (in bold) are separated by an intron (italics). In the transcriptome (above), the splicing together of these exons gives rise to reads $T=\{\texttt{CACAT}, \texttt{TCACA}\}$ containing the exon/exon boundary whereas, in the genome, the reads $G=\{\texttt{AGACA}, \texttt{GACAT}\}$ extend from the \texttt{ACAT} exon into the intron.
}\label{Fig1}
\end{figure}

Figure~\ref{Fig1} shows a very simple example of how splicing in the transcriptome might give rise to a read set $T$ containing sequence that is not present in the genomic reads $G$. Figure~\ref{Fig2} shows the BWTs of the two datasets. In the function \texttt{doSharedBehaviour()}, we look for $Q$-intervals shared by $\mathrm{BWT}(T)$ and $\mathrm{BWT}(G)$ for which the $Q$-interval in $\mathrm{BWT}(G)$ has a unique backward extension but the corresponding $Q$-interval in  $\mathrm{BWT}(T)$ exhibits significant evidence of one or more different backward extensions $cQ$. In our implementation, spurious junction predictions due to sequencing error are guarded against by ignoring any such backward extensions for which the number of occurrences (given by the number of $c$ symbols present in the $Q$-interval) fails to exceed a threshold. Any $T$-only $cQ$-intervals that do pass this test are backward-extended in subsequent intervals by \texttt{doAOnlyBehaviour()} until a string $CQ$ is obtained for which the size of the $CQ$-interval fails to exceed a threshold $t$, which is equivalent to demanding that $CQ$ must occur at least $t$ times in $T$. The aim of this extension is to accumulate as much sequence context as possible to the left of the putative exon/exon junction. In a similar way, we could improve specificity by allowing \texttt{doBOnlyBehaviour()} to extend $G$-only intervals and thus accumulate sequence context that reaches into the separating intron, although our current implementation does not do this.

If the sequence context to the right of a predicted junction is a prefix of the sequence that lies to the right of another prediction junction, then the former prediction is subsumed into the latter. For example, in Figure~\ref{Fig1} the same splice junction gives rise to reads \texttt{TCACA} and \texttt{CACAT} with rightward contexts \texttt{ACA} and \texttt{ACAT}: the former is a prefix of the latter. This aggregation of predictions is conveniently done by making a single pass through the final list of predictions once they have been sorting  in lexicographic order of their rightmost context. As well as removing repeated predictions, this acts as a further guard against false positives - we discard any predictions whose contexts cannot be rightward-extended in this way. 

Finally, we note that the double-stranded nature of DNA is handled by aggregating the individual chromosomal sequences plus their reverse complements into a sequence collection and building the BWT of that.

\begin{figure}[!htb]
{\scriptsize
  $$
    \begin{array}[t]{cl}
    $\textrm{BWT}(T)$ & \textrm{suffixes} \\
    \texttt{T}  & \texttt{$\$_1$} \\
    \texttt{A}  & \texttt{$\$_2$} \\
    \texttt{C}  & \texttt{A$\$_2$} \\   
    \texttt{C}  & \texttt{ACA$\$_2$} \\
    \texttt{C}  & \texttt{ACAT$\$_1$} \\
    \texttt{C}  & \texttt{AT$\$_1$} \\
    \texttt{A}  & \texttt{CA$\$_2$} \\
    \texttt{T}  & \texttt{CACA$\$_2$} \\
    \texttt{$\$_1$}  & \texttt{CACAT$\$_1$} \\
    \texttt{A}  & \texttt{CAT$\$_1$} \\
    \texttt{T}  & \texttt{T$\$_1$} \\
    \texttt{$\$_2$}  & \texttt{TCACA$\$_2$} \\

    \end{array}%
    \qquad
  \begin{array}[t]{cl}
    $\textrm{BWT}(G)$ & \textrm{suffixes} \\

    \texttt{A}  & \texttt{$\$_1$} \\
    \texttt{T}  & \texttt{$\$_2$} \\
    \texttt{C}  & \texttt{A$\$_1$} \\
    \texttt{G}  & \texttt{ACA$\$_2$} \\
    \texttt{G}  & \texttt{ACAT$\$_2$} \\
    \texttt{$\$_1$}  & \texttt{AGACA$\$_1$} \\
    \texttt{C}  & \texttt{AT$\$_2$} \\
    \texttt{A}  & \texttt{CA$\$_1$} \\
    \texttt{A}  & \texttt{CAT$\$_2$} \\
    \texttt{A}  & \texttt{GACA$\$_1$} \\
    \texttt{$\$_2$}  & \texttt{GACAT$\$_2$} \\
    \texttt{A}  & \texttt{T$\$_2$} \\

  \end{array}%
    $$
}

\caption{
Comparison of the BWTs of $T$ and $G$ from Figure~\ref{Fig1}. During the second execution of the procedure in Figure~\ref{Alg2}, we find the  \texttt{AC}-interval in $\textrm{BWT}(G)$ has a unique backward extension \texttt{G}, but the corresponding interval is $\textrm{BWT}(T)$ has a different backward extension \texttt{C}. This is corroborated (and the sequence context to the right of the splice junction is extended) at steps 3 and 4 when the \texttt{ACA}- and \texttt{ACAT}-intervals of the two BWTs are compared. The \texttt{CA}-interval also suggests a divergent backward extension, but the lack of a forward extension of \texttt{CA} that is common to both $T$ and $G$ means this observation is not corroborated by subsequent executions of the code in Figure~\ref{Alg2} and is therefore discarded.
}\label{Fig2}
\end{figure}

\section{Results}

\subsection{Reference-free detection of splice junctions}

We tested our approach using data from a recent study~\cite{Murchison2012} during which $1.45$ billion genomic reads and $132$ million RNA-Seq reads, all $100$bp in length, were obtained from an individual Tasmanian devil. The RNA-Seq library was prepared from a mixture of mRNA from 11 different tissues to obtain broad coverage of gene content. The genome of the Tasmanian devil was estimated to be between $2.89$ and $3.17$ gigabase pairs (Gb) in size and is thus comparable in size to the human genome. \emph{De novo} assembly of the genomic reads yielded $3.17$ Gb of sequence with an N50 of $1.85$ megabase pairs (Mb). The Ensembl gene annotation pipeline was then applied to the assembled contigs: evidence from alignment of mammalian EST, protein and RNA-Seq sequences was combined and then various gene prediction algorithms were used to refine these alignments and to build gene models (more detail on the annotation procedure is given in the supplement of~\cite{Murchison2012}). We obtained the most recent version ($0.67$) of the Tasmanian devil gene annotation from the Ensembl FTP site. It contains \numprint{20456} genes which give rise to \numprint{187840} exon 
junction sites.

We built BWTs of both the genomic and RNA-Seq read sets using the algorithms given in \cite{BauerCoxRosoneCPM11} and compared them as described in the previous sections. The sequences to the right and left of each prediction were aligned to the devil assembly using BWA \cite{bwa2009} in single-read mode, setting the option to allow up to $10$ candidates for each read. Predictions for which the left and right halves aligned to the same contig with appropriate orientation were classified as putative junction sites: we obtained \numprint{171371} of these.

We also predicted gene models and junction sites from the same RNA-Seq reads using version 2.0.0 of Tophat \cite{Trapnell2009}, which is a popular tool for this task. Tophat first aligns reads to a reference genome with the Bowtie2 aligner~\cite{bowtie2009} then builds splicing models based on 
these alignments. 
The results of our comparison are summarized in Table~\ref{TabResults}: Tophat predicts \numprint{120010} junction sites, of which \numprint{66587} are not contained in the gene annotation.

\begin{figure}[!htb]
\begin{tabular}{c|c|c|c|c|c}
Tool & Junctions predicted & True positives & False Negatives & Sensitivity (\%) & FDR (\%) \\
\hline
BWT & \numprint{171371} & \numprint{93615} & \numprint{94225} & $49.84$ & $45.37$ \\
Tophat & \numprint{120010} & \numprint{66587} & \numprint{121253} & $35.45$ & $44.51$  \\
\end{tabular}
\caption{Comparison of junction site predictions. Our approach predicts \numprint{171371} sites and Tophat predicts \numprint{120010}. Treating the Ensembl annotation
as a gold standard, we evaluate sensitivity and false discovery rate of each method. The BWT-based approach is competitive with the established
software Tophat.}\label{TabResults}
\end{figure}

Using the BEDtools software suite~\cite{Quinlan2010}, we identified junction sites that overlap with sites contained in the Ensembl 
gene annotation. We used default parameters, apart from requiring a reciprocal overlap of $90$\% of the feature length.
Of the \numprint{171371} sites computed by our approach, \numprint{94225} match known Ensembl predictions. Manual inspection of the remaining sites revealed that 
many were contained in putative gene annotation derived from EST alignments or from \emph{ab initio} gene recognition algorithms. These putative annotations were not incorporated into the final annotation because of various threshold or partially contradicting evidence. They represent nevertheless likely candidates for coding regions. The EST alignments cover $48.06$~Mb in \numprint{22582} alignments and
the \emph{ab initio} predictions cover $44.92$~Mb in \numprint{44659} regions. 
Of the \numprint{77756} junction sites detected by our method that did not have a counterpart in the Ensembl prediction, \numprint{24322} did not have a 
match in the EST alignment 
data set and \numprint{11168} did not match coding regions predicted by \emph{ab initio} algorithms. Taking these sets together, we found that only \numprint{8755} out of \numprint{171371} (5.11\%) did not have any evidence of being in transcribed regions. For Tophat, \numprint{53423} junction sites did not have a match in the 
Ensembl gene annotation. Of these predictions, \numprint{14668} did not have a match in regions covered by EST alignments and \numprint{6227} did not have a match in \emph{ab initio} gene predictions. In sum, \numprint{4732} Tophat predictions (3.94\%) did not match any potentially coding regions.

\section{Discussion}

In this work, we show that BWTs of transcriptomic and genomic
read sets can be compared to obtain reference-free predictions of splice junctions
that have high overlap with results from more standard reference-based methods.
Our method predicts splice junctions by directly comparing sets of genomic and
transcriptomic reads and can therefore provide orthogonal confirmation of gene
predictions obtained by comparative genomics approaches. A reference sequence
is not required for the prediction process itself (here we map the predicted junctions to the assembly only for comparison purposes), making the method particularly well suited to the analysis of organisms for which no reference genome
exists.

When comparing the performance of our method with Tophat we find that, at least on this data, our approach has superior sensitivity and comparable false discovery rate. In order to give a strong proof of principle we deliberately avoided building any sort of prior information about gene structure into our analysis. In contrast, Tophat makes assumptions about the presence of canonical dinucleotide motifs at donor/acceptor sites and the relative abundance of isoforms. This is an entirely reasonable thing to do, but it is conceivable that Tophat's use of prior information might be a disadvantage for this particular dataset as it is not clear to what extent these signals are conserved across species and in particular in the Tasmanian devil. 

An obvious piece of prior information needed by both Tophat and the Ensembl annotation pipeline is of course a reference sequence. Although considered to be of `draft' quality, the Tasmanian devil assembly we used \cite{Murchison2012} nevertheless required not only both considerable computational and manual effort to generate but also made use of additional sequencing data in the form of long-insert mate pair libraries.

 While the Ensembl annotation pipeline is a robust and well-established methodology, we note that our implicit treatment of the Ensembl annotation as absolute truth is an assumption that might be questioned, since the pipeline is being applied here to a draft assembly from a relatively poorly-understood genome. Nevertheless, we believe that our results do demonstrate that direct comparison of BWTs gives results that are biologically credible and that are competitive with existing tools.

The need to sequence the genome as well as the transcriptome means our method is unlikely to supplant methods such as Tophat which can operate on transcriptome data alone. Comparison of transcriptome to exome data might be more practical and have some utility, although it is arguable whether such an approach remains hypothesis-free. However in situations where, as here, both genome and transcriptome data are available our method may provide valuable additional information. Even for a much better characterized genome such as human, our algorithm should provide insight into transcription from regions that are not well-represented in the reference sequence and might also be a useful tool for investigating RNA editing. A further improvement of the method would be to used read-pairing information to link junction sites that are present in the same read pair and hence in the same transcript.

Computing the BWTs of the genomic and transcriptome read sets took around 6 days and 12 hours of wallclock time respectively, although the method employed ran entirely in external memory and so did not require high-end computing hardware.  Moreover, our previous work \cite{BauerCoxRosoneCPM11} suggests that these computation times could be approximately halved by storing the work files on a SSD (flash memory) drive and could be further improved by using a different algorithm that reduces I/O at the expense of moderate RAM usage. Indeed, one could make a case that the cost of BWT computation should not be included in the overall compute time, since it is useful in its own right for lossless compression of the data \cite{Cox2012} and for facilitating other analyses such as de novo assembly \cite{Simpson2010,Simpson2011}.

The comparison of BWTs ran in just under $3$ days of wallclock time. Again, all processing was done in external memory and could be sped up by the use of an SSD drive or, alternatively, the sequential nature of the algorithm's I/O access would facilitate cache-efficient processing if the BWT files were instead held in RAM on a high-end machine. To put these numbers into context, the analysis using TopHat took
$18.6$ hours but this obviously does not include the time to assemble and curate the reference genome.

In addition, our implementation is a proof-of-principle with considerable scope for optimization. Future work will focus on such improvements and on exploring further applications of the algorithm described in Figure~\ref{Alg2}: many important tasks in sequence analysis can be reinterpreted as a comparison between BWTs, not least the comparison of tumour and normal read sets from cancer samples and the comparison of reads to a reference sequence.

\section*{Acknowledgement}

A.J.C. and O.S.-T. are employees of Illumina Inc., a public company that develops and markets systems for genetic analysis, and receive shares as part of their compensation. Part of T.J.'s contribution was made while on a paid internship at Illumina's offices in Cambridge, UK.
We thank Elizabeth Murchison and Zemin Ning for contributing the genomic and RNA-Seq data and the genome assembly of the Tasmanian Devil.

\bibliographystyle{plain}
\bibliography{BWT}

\end{document}